# Noise propagation and MP-PCA image denoising for high-resolution quantitative $T_2^*$ and magnetic susceptibility mapping (QSM)

Liad Doniza[1], Mitchel Lee[2], Tamar Blumenfeld Katzir[3], Moran Artzi[4,5,6], Dafna Ben Bashat[4,5,6], Dvir Radunsky[3], Karin Shmueli[2], Noam Ben-Eliezer[3,5,7]

[1] Department of Electrical Engineering, Tel Aviv University, Tel Aviv, Israel, [2] Department of Medical Physics and Biomedical Engineering, University College London, London, UK, [3] Department of Biomedical Engineering, Tel Aviv University, Tel Aviv, Israel, [4] Sagol Brain Institute, Tel Aviv Medical Center, Tel Aviv, Israel, [5] Sagol School of Neuroscience, Tel Aviv University, Tel-Aviv, Israel, [6] Sackler Faculty of Medicine, Tel Aviv University, Tel Aviv, Israel, [7] Center for Advanced Imaging Innovation and Research (CAI2R), New-York University Langone Medical Center, New York, NY, United States

*Abstract*— **Quantitative Susceptibility Mapping (QSM) is a technique for measuring magnetic susceptibility of tissues, aiding in the detection of pathologies like traumatic brain injury and multiple sclerosis by analyzing variations in substances such as iron and calcium. Despite its clinical value, achieving high-resolution QSM (voxel sizes < 1 mm$^3$) reduces signal-to-noise ratio (SNR), compromising diagnostic quality. To mitigate this, we applied the Marchenko-Pastur Principal Component Analysis (MP-PCA) denoising technique on $T_2^*$ weighted data, to enhance the quality of $R_2^*$, $T_2^*$, and QSM maps. Denoising was tested on a numerical phantom, healthy subjects, and patients with brain metastases and sickle cell disease, demonstrating effective and robust improvements across different scan settings. Further analysis examined noise propagation in $R_2^*$ and $T_2^*$ values, revealing lower noise-related variations in $R_2^*$ values compared to $T_2^*$ values which tended to be overestimated due to noise. Reduced variability was observed in QSM values post denoising, demonstrating MP-PCA's potential to improve the diagnostic value of high resolution QSM maps.**

## I. INTRODUCTION

### A. Motivation for high-resolution QSM

Quantitative susceptibility mapping (QSM)[1], [2] is a common MRI technique for measuring the magnetic susceptibility of tissues ($\chi$). QSM is mainly utilized for detecting pathologies that involve dysregulations in iron, calcium, and myelin content, and is routinely used as a biomarker for pathological conditions such as microbleeds in traumatic brain injury, sickle cell disease, brain metastases, glioma, multiple sclerosis, and Parkinson's disease [2]–[9]. Several works [2], [9]–[11] have shown that increasing the spatial resolution of clinical data lowers partial volume effects and offers more faithful mapping of QSM values, e.g., of small bin structures. Specifically, high-resolution susceptibility maps are important for separating subthalamic nuclei, and characterization of the internal structure of the substantia nigra (SN), which plays a central role in Parkinson's disease and in planning deep brain stimulation surgeries [12]–[14]. Additionally, QSM is useful for detecting paramagnetic rims of iron-laden active microglia and macrophages in multiple sclerosis (**MS**) lesions, correlating with cognitive function and with the expanded disability status scale (**EDSS**) [15], [16]. Collecting high resolution data, however, comes with a tradeoff of lower SNR, and reduced diagnostic quality [10], [11].

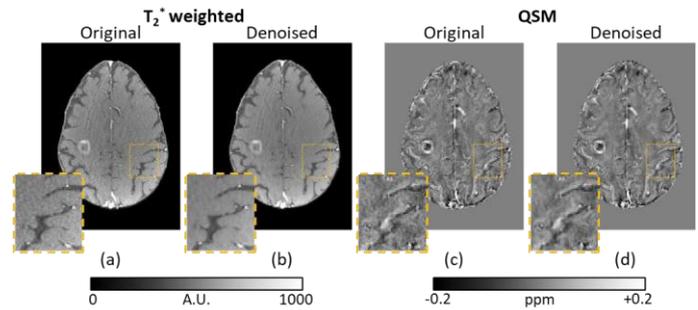

**Fig. 1.** The original and denoised acquired images along with their corresponding quantitative susceptibility maps. (a) shows the original magnitude image, (b) presents the denoised image, and (c-d) display their respective QSM maps. A zoomed-in region of each image is shown at the bottom left corner.

Typically, QSM values are reconstructed from complex $T_2^*$ weighted ($T_2^*w$) images, acquired using a 3D multi gradient echo (**GRE**) protocol. Due to the intricate processing pipeline and the use of complex data, noise propagation affects QSM values beyond simple reduction in SNR, and may lead to artifacts such as streaking, or image distortions due to inaccurate phase unwrapping [2], [9]. Existing solutions to this problem are based on adding regularizations or adjusting hyperparameters in the QSM processing pipeline [17], [18]. These methods, however, rely on a priori knowledge of the magnetic susceptibility of the tissue, which is often unavailable and varies across the imaged organs or in the presence of pathologies.

### B. Denoising QSM - What has already been done

An effective approach to avoid noise related artifacts and produce high-resolution QSM maps is to denoise the raw $T_2^*w$ data that is used as input to the QSM pipeline. Bazin *et al.* [19] implemented a technique that employs Principal Component Analysis (**PCA**) to separate signal and noise in magnetization-prepared gradient echo sequence (**MPRAGE**) data, acquired at 7T. Zhang *et al.* [20] utilized a self-supervised deep neural network for denoising of echo planar imaging (**EPI**) data, while utilizing data from multiple slices and echo-times [21]. This model, however, is designed to denoise only the magnitude, rather than complex images, and may therefore produce less optimal results.

Some approaches, such as Plug-and-Play QSM [22], and DeepQSM [23] or QSMnet [24] supervised deep learning

models, have been developed to denoise complex single or multi-GRE data. These methods operate within the QSM pipeline, after the field inhomogeneity map has been generated from noisy data, which can inadvertently introduce inherent noise, potentially impacting the results. Furthermore, deep learning approaches may encounter limitations in terms of generalization. These networks require a diverse training dataset, including variations in resolution or susceptibility $\chi$, to function effectively. A lack of such diversity in the training data can lead to suboptimal reconstruction performance [23]–[26].

### C. Existing applications of MP-PCA denoising

Recent studies have demonstrated a significant denoising effect by applying the Marchenko-Pastur (MP) criteria in PCA for effective noise reduction. This method relies on the fact that for MRI signals, the eigenvalues of noise obey the universal Marchenko-Pastur law of random matrix theory [27]. Its applications have been explored in various MRI modalities, including Diffusion-weighted MRI [28], enhancement of parameter estimation in multiexponential relaxometry [29], production of more precise $T_2$ maps [30], multi-modal acquisition for quantitative mapping of the spinal cord [31], language mapping in functional MRI for patients with brain tumors [32], and diffusion-weighted spectroscopy [33].

In this work, MP-PCA was employed for denoising multi-GRE data, which is both utilized and recommended for QSM calculations [34]. MP-PCA was integrated as an initial step in processing $T_2^*w$ images within the QSM pipeline for brain imaging [35] (Fig. 1), thereby, also assessing improvements in both $T_2^*w$ images and $T_2^*$ maps. This novel denoising pipeline was successfully applied to a numerical phantom and in vivo brain data, covering healthy individuals, and patients with sickle cell disease and brain metastases. These applications were conducted across a range of resolutions and scan parameters, highlighting the method's robustness.

## II. THEORY

### A. Theory of MP-PCA denoising

For readers' convenience, we provide a concise overview of MP-PCA denoising theory, as applied to $T_2^*w$ imaging. A more comprehensive description can be found in [28]–[30]. The $T_2^*w$ magnitude and phase data are acquired using a 3D GRE sequence. We denote the acquired 4D complex data matrix as $D \in \mathbb{C}^{[N_x, N_y, N_z, N_E]}$, where three dimensions represent spatial domain, and the fourth denotes the number of echoes $N_E$. A sliding 4D window denoted as $W \in \mathbb{C}^{[n_x, n_y, n_z, N_E]}$ is used to traverse the entire dataset $A$, while denoising is applied within that window. A natural choice for 3D dataset with isotropic resolution is a 3D isotropic window size of $N_W \times N_W \times N_W$. The matrix $W$ in each voxel is vectorized according to $N_V = N_W^3$ into a 2D matrix $A \in \mathbb{C}^{N_V \times N_E}$. $A$ is then normalized to $\hat{A}$ by subtracting the mean signal within each 3D spatial window $\bar{A} \in \mathbb{C}^{1 \times N_E}$ for each echo time (**TE**), expressed as:

$$\hat{A}(i) = A(i) - \bar{A}(i) \cdot 1_{N_V \times 1}\ ; \ \forall i \in 1 \dots N_E \quad (1)$$

$\hat{A}$ is then deconvolved into its linearly independent sources or principal components (**PCs**) using singular value decomposition (**SVD**) [36].

$$SVD(\hat{A}) = U\Lambda V^T \quad (2)$$

Here, the diagonal $M = \min(N_V, N_E)$ elements of $\Lambda$ represent the singular values of $\hat{A}$, and $\Lambda^2$ is an $M \times M$ diagonal matrix containing the eigenvalues $\lambda_1 \dots \lambda_M$. For each $\hat{A}$ the first $P$ eigenvalues can be associated with the signal, whereas the remaining $M - P$ represent noise [27]–[29], under the assumption that the noise is characterized by Marchenko-Pastur distribution. The value of $P$ is determined by the minimum value of P which holds the following inequality [28]:

$$\frac{\sum_{i=P+1}^{M} \lambda_i}{M - P} > \frac{\lambda_{P+1} - \lambda_M}{4\sqrt{\frac{M-P}{N_V}}} \quad (3)$$

Noise is then removed by truncating the last $\lambda_{P+1} \dots \lambda_M$ eigenvalues and transforming back to the spatial domain using:

$$\hat{A}' = U\Lambda' V^T \quad (4)$$

Finally, the mean value at each time point $\bar{A}$ is added back to $\hat{A}'$. This process iterates across the entire spatial domain, moving $W$ one voxel at a time. The denoised signal at each voxel is then computed as the average value, produced by all windows that include that voxel [37].

### B. Quantitative mapping of $R_2^*$ and $T_2^*$ relaxation

In order to estimate the $R_2^*$ relaxation of the tissue and its inverse, the $T_2^*$, an exponential fitting for each voxel of the time series is applied [38]. The calculation was done to examine the noise propagation in the fitting.

### C. Quantitative mapping of magnetic susceptibility

QSM maps were computed from magnitude and phase images. Signal phase is determined by the magnetic field inhomogeneities, $\Delta B_0(\vec{r})$ which, in turn, represent a convolution of the tissue susceptibility $\chi(\vec{r})$ with the dipole field distribution $d(\vec{r})$ [2]:

$$\Delta B_0(\vec{r}) = B_0 \cdot \{\chi(\vec{r}) * d(\vec{r})\} \quad (5)$$

This equation known as the forward problem, where $B_0$ is the static homogeneous magnetic field in the z-direction, and $d(\vec{r})$ is defined as:

$$d(\vec{r}) = \frac{3\cos^2\varphi - 1}{4\pi\vec{r}^3} \quad (6)$$

Here, $\varphi$ is the angle between $\vec{r}$ and the magnetic field's direction.

The reconstruction of QSM can be done with several pipelines and methods. In our study the maps were calculated using the pipeline described by Karsa et al. [35]. This process begins with non-linear fitting of the complex time-series data [39]–[42] to calculate the field map $\Delta B_0(\vec{r})$ and a weighting matrix $W_N$. This weighting matrix is proportional to the magnitude images along multiple echoes, and used to compensate for variations in noise in $\Delta B_0(\vec{r})$ as described in [40]. This step was followed by Laplacian phase unwrapping [42], [43]. The subsequent step involved generating a brain mask using FSL's Brain Extraction Tool (**BET**) [44], combined with a mask based on thresholding $W_N$ at its mean.

Projection onto Dipole Fields (**PDF**) [45], derived from the projection theorem of Hilbert space [46], was then used to remove background fields induced by dipoles outside of the region of interest (**ROI**), thus isolating the field due to tissue-related dipoles, termed the local magnetic field.

Retrieving the susceptibility maps from the field map is known as the inverse problem, for which the solution was implemented using iterative least-square fitting in the image domain with Tikhonov regularization [40], [47]:

$$\underset{\chi}{\mathrm{argmin}} \| M \cdot W_N (B_L - B_0 \cdot (d * \chi)) \|_2^2 + \alpha \| \chi \|_2^2 \quad (7)$$

where $B_L$ represent the local magnetic field, and $\alpha$ denotes Tikhonov regularization.

## III. METHODS

### A. Validations on a numerical phantom

To assess the efficiency and limitations of the MP-PCA denoising process we constructed a numerical phantom, designed as a cylinder aligned along the z-axis (the direction of $B_0$) and filled with four smaller cylindrical tubes with varying magnetic susceptibilities. Each tube's proton density was set to 1 $[a.u.]$. In order to assess the denoising quality at realistic condition the simulation parameters were matched to typical experimental parameters using an isotropic resolution of $0.75\ mm^3$ and $B_0 = 3\ T$. $\Delta B_0(\vec{r})$ was calculated using the forward model in (5) and depended only on differences in magnetic susceptibility, meaning that background (i.e., hardware related) field variations were not included in the simulation. The phase within each voxel was then derived based on:

$$\theta(\vec{r}) = 2\pi \cdot \gamma \cdot (B_0 + \Delta B_0(\vec{r})) \cdot TE \quad (8)$$

Here $\gamma$ is the hydrogen gyromagnetic ratio and $TE$ is the echo time.

Next, we calculated the transverse relaxation values of the phantom. $T_2$ values in each voxel were calculated based on magnetic susceptibility and relaxivity $\mathcal{R}$ [s$^{-1}$·mM$^{-1}$] of Gd-DTPA, a common contrast agent used in clinical imaging [48], [49]. $T_2'$, which corresponds to field inhomogeneity caused by the susceptibility variations, was calculated as:

$$T_2' = (2\pi \cdot \gamma \cdot \Delta B_0(\vec{r}))^{-1} \quad (9)$$

Using these two values, we set the $T_2^*$ value at each voxel according to the known relation: $1/T_2^* = 1/T_2 + 1/T_2'$.

Lastly, a series of $T_2^*w$ images were calculated for $N_E = 8$ echo times, with the following TEs: $TE_1 = 3\ ms$; $\Delta TE = 4\ ms$. Complex white gaussian noise was added to the series of $T_2^*$-weighted images using MATLAB's *wgn* function (The MathWorks, Natick, MA) at SNRs of 10 and 20 with respect to the highest magnitude of the $T_2^*w$ image. This process was repeated 16 times, with different randomized noise patterns. Denoising was applied on each image separately using pipeline described in Section II above, and a $2 \times 2 \times 2 \times N_E$ window.

$R_2^*, T_2^*$ and values were calculated using exponential fitting of the time series, and QSM values were calculated the QSM pipeline pre- and post-denoising. Mean and standard deviation (**SD**) of $R_2^*, T_2^*$ and QSM values were calculated per tube, for the first repetition. A goodness-of-fit of the exponential $R_2^*$ fitting was calculated for each voxel post-denoising, and voxels with $R^2 < 0.8$ were excluded from the calculation of the $R_2^*$ and $T_2^*$ mean and SD values. The same voxels were also excluded from the mean and SD, calculated for the original $R_2^*, T_2^*$ maps.

SNR maps were calculated for $T_2^*w$, quantitative $T_2^*$ maps, and QSM maps from the 16 repetitions, according to:

$$SNR_{map}(\vec{r}) = \frac{\bar{I}(\vec{r})}{\sigma(\vec{r})} \quad (10)$$

where $\bar{I}$ and $\sigma$ are the mean and SD of the signal in each voxel across 16 repetitions. Average SNR was estimated for each tube based on the mean SNR across all voxels in the tube.

### B. Application on healthy volunteers and patients with brain metastases and sickle cell disease

#### 1) MRI Scans

Four healthy volunteers and two patients were scanned using 3D multi-GRE protocols (see TABLE I for detailed scan parameters). The volunteers and two patients with brain metastases were scanned on a 3T Siemens Prisma scanner. The third patient had sickle cell disease and was scanned on a 1.5T Philips Achieva scanner. Informed written and verbal consent were obtained for all participants. Healthy volunteers' scans, scans of two volunteers with brain metastases, and scans of volunteer with sickle cell disease were approved by the local institutional review board (**IRB**) committee.

TABLE I
SCAN PARAMETERS, MULTI-GRE ACQUISITIONS.

| Scan No. | State | Voxel size [mm$^3$] | TR [ms] | TE$_1$ [ms] | ΔTE [ms] | No. of Echoes | FA [deg] |
|---|---|---|---|---|---|---|---|
| 1 | Healthy | 0.6 | 45 | 3.99 | 5.24 | 8 | 15 |
| 2 | | 0.75 | | | | | |
| 3 | Healthy | 0.5 | 68 | 8.00 | 8.00 | 8 | 15 |
| 4 | | | | | | | |
| 5 | Healthy | 0.75 | 37 | 2.99 | 4.25 | 8 | 12 |
| 6 | | 1 | | | | | |
| 7 | Healthy | 0.6 | 37 | 2.99 | 4.25 | 8 | 12 |
| 8 | | 0.75 | | | | | |
| 9, 10 | Brain Metastasis | 0.75 | 45 | 3.99 | 5.24 | 8 | 15 |
| 11 | Sickle cell disease | 1.5 | 27.4 | 4.28 | 4.94 | 5 | 15 |

Healthy volunteers were also scanned using an MP2RAGE protocol for brain segmentation, with the following parameters: slice thickness = $1\ mm$, matrix size = $192 \times 156$, field of view (**FOV**) = $192 \times 156\ mm^2$, TE/Repetition time (**TR**) = $3.52/4000\ ms$, $N_{slices} = 192$, Grappa acceleration factor = 2, total scan time = 6:00 min.

#### 2) Data processing

Denoising was applied on $T_2^*w$ complex DICOM images from all scans using a window size of $W = 2 \times 2 \times 2 \times N_E$, followed by generating quantitative $T_2^*$ and QSM maps. MP2RAGE images were segmented and then registered to the multi-GRE image space using Freesurfer software [50]. Denoising was assessed in three regions of interest (**ROIs**) including globus pallidus, caudate nucleus, and putamen (contralateral regions were analyzed separately). An erosion of one pixel was applied to all ROIs to avoid partial volume effects.

Mean and SD of $T_2^*w$ values, $T_2^*$ maps, and QSM values were estimated in each ROI for a single representative slice. SNR of $T_2^*w$ data was assessed on magnitude images by dividing the mean signal in each ROI by the SD of four background rectangles similar to [30], [51]. Noise in magnitude images obey the Rayleigh distribution [52], whereas the denoising process was applied on complex data, where noise is characterized by Gaussian distribution. To correct for this difference, SD values of background voxels were divided by the factor $\sqrt{2-\pi/2}$ [51].

## IV. RESULTS

$T_2^*w$ images, and QSM maps of the numerical phantom are illustrated in Fig. 2 for SNRs of 10 and 20 pre- and post-denoising. Fig. 3 shows the corresponding $R_2^*$, and $T_2^*$ maps. The denoising procedure enhanced the quality of all images and maps, producing more homogeneous structures, which are closer to ground truth values. Quantitative values, corresponding to each tube, are detailed in TABLE II. Seeing as the simulated noise was generated with zero mean, its addition and removal did not change the baseline values of the real and imaginary channels of the $T_2^*w$ complex dataset, also indicating that no bias was introduced by the denoising process. SD of the $T_2^*w$ images values decreased by $76.6 \pm 0.0$ % post-denoising (averaged across four tubes), corresponding to an increase of $324.2 \pm 1.0$ % in SNR. Fitting of $R_2^*$ values produced an error of $0.2 \pm 2.4$ % pre-denoising, which decreased to $0.0 \pm 0.1$ % post-denoising. The same effect was observed in the SD of $R_2^*$ values, which exhibited an average decrease of $75.4 \pm 0.9$ % post denoising, corresponding to an increase of $321.0 \pm 7.0$ % in SNR. Quantitative $T_2^*$ values were overestimated by $35.8 \pm 28.0$ % pre-denoising, which was removed post denoising, reflected as a negligible deviation of $0.8 \pm 0.4$ % vs. original $T_2^*$ values. SD of $T_2^*$ values decreased by $91.3 \pm 5.2$ % post denoising, corresponding to an increase of $459.2 \pm 51.6$ % in SNR.

Notably, although $T_2^*$ is the inverse of $R_2^*$, a drastically different bias emerged between the two types of values. This was reflected in an average increase of 72.2 % in the coefficient of variation (**CV**) of $T_2^*$ compared to $R_2^*$. This can be attributed to noise propagation during the transition of $R_2^*$ to $T_2^*$ values, which involves a division operation ($T_2^* = 1/R_2^*$). This changes the noise distribution, leading to a strong overestimation of $T_2^*$ values. The effect was more pronounced for higher $T_2^*$ values, as they correspond to $R_2^*$ values that are closer to zero.

QSM maps demonstrated no bias in mean values both pre- and post-denoising. We ascribe this stability to the QSM pipeline's reliance on a spatial magnitude weighting matrix $W_N$, as reported in [53]. As expected, the SD in QSM values was smaller by $30.9 \pm 3.5$ % post-denoising, corresponding to an increase of $304 \pm 3.8$ % in SNR. An example of the effect of denoising on the raw signal decay curves is shown in Fig. 4, juxtaposing the original, pre-denoising, and post-denoising $T_2^*w$ signals. Denoised signals (red '*') exhibit significantly reduced variability compared to the noisy signals (blue '■'), closely following the original decay curves (green '+').

$T_2^*w$ images, $T_2^*$ maps, and QSM maps for a healthy volunteer are shown in Fig. 5 pre- and post-denoising. The denoising process did not introduce any visible blurring, preserving the fine anatomical details, as exemplified in the zoomed-in insets (middle and bottom rows). Quantitative of $R_2^*$, $T_2^*$, and QSM values are delineated in TABLE III for each ROI. An average increase of $1.1 \pm 0.6$ % in the $R_2^*$ values was observed post denoising across all ROIs, with a corresponding decrease of $20.3 \pm 7.0$ % in SD.

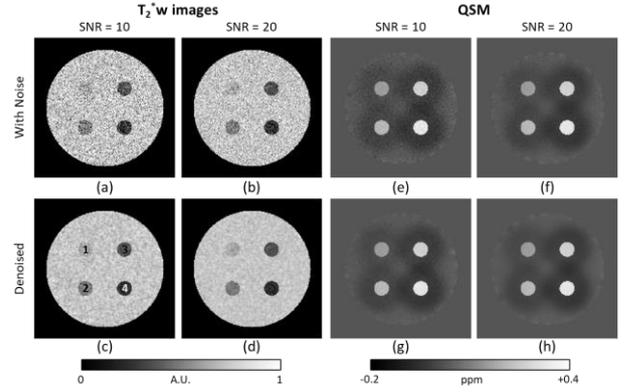

**Fig. 2.** MP-PCA denoising of a numerical phantom. (a-d) $T_2^*w$ images (4$^{th}$ echo) pre- and post-denoising. (e-h) QSM maps pre- and post-denoising. Simulations were done at low SNRs of 10 & 20. (a) shows the location of tubes (1)-(4).

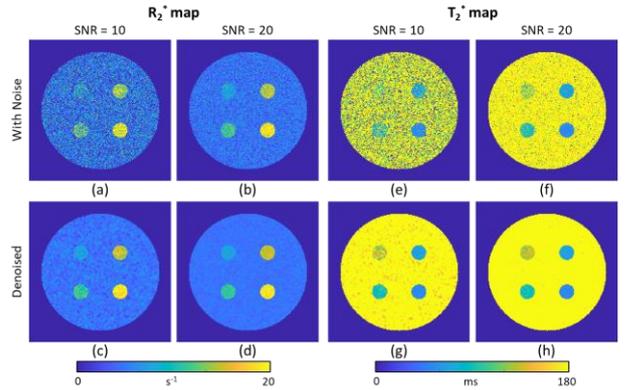

**Fig. 3.** The corresponding $R_2^*$ and $T_2^*$ values for Fig. 2 MP-PCA denoising of the numerical phantom at SNRs of 10 & 20. (a-d) $R_2^*$ maps pre- and post-denoising. (e-h) $T_2^*$ maps pre- and post-denoising.

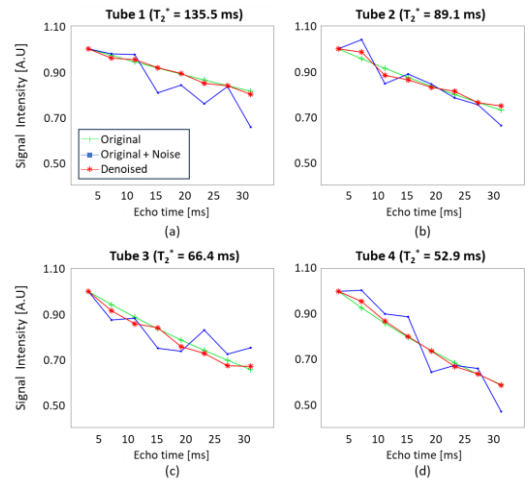

**Fig. 4.** $T_2^*$ decay in all four tubes in the numerical simulation. The green line ('+' marker) denotes the original decay curve, the blue ('·' marker) denotes the original decay curve, added with noise, and the red line ('*' marker) denotes the denoised signal. Efficient denoising is achieved for all three data types and two SNRs without visible loss of information.

TABLE II
MEAN, SD, AND SNR OF $T_2^*w$, $R_2^*$, $T_2^*$, AND QSM VALUES, CALCULATED FOR EACH TUBE IN THE NUMERICAL PHANTOM. ALL VALUES WERE CALCULATED FROM 16 REPETITIONS.

| Tube No. | $T_2^*w$ values (1st echo) | | | | | | $R_2^*$ map | | | | | | $T_2^*$ map | | | | | | QSM | | | | | |
|---|---|---|---|---|---|---|---|---|---|---|---|---|---|---|---|---|---|---|---|---|---|---|---|---|
| | Mean [a.u] | | SD [a.u] | | SNR | | Mean [s⁻¹] | | SD [s⁻¹] | | SNR | | Mean [ms] | | SD [ms] | | SNR | | Mean [ppb] | | SD [ppb] | | SNR | |
| | Org. | w/ noise | Den. | w/ noise | Den. | w/ noise | Den. | Org. | w/ noise | Den. | w/ noise | Den. | w/ noise | Den. | Org. | w/ noise | Den. | w/ noise | Den. | w/ noise | Den. | Org. | w/ noise | Den. | w/ noise | Den. | w/ noise | Den. |
| #1 | 0.98 | 0.98 | 0.98 | 0.10 | 0.02 | 10.5 | 44.5 | 7.4 | 7.7 | 7.4 | 3.8 | 1.0 | 1.9 | 7.6 | 135.5 | 247.2 | 137.3 | 792.0 | 18.8 | 1.2 | 7.4 | 148.3 | 148.5 | 148.5 | 17.9 | 29.1 | 18.43 | 6.7 | 27.2 |
| #2 | 0.97 | 0.97 | 0.97 | 0.10 | 0.02 | 10.4 | 44.3 | 11.2 | 11.2 | 11.2 | 4.4 | 1.1 | 2.6 | 10.8 | 89.1 | 117.9 | 89.8 | 128.6 | 8.6 | 1.8 | 10.7 | 208.6 | 209.1 | 209.2 | 22.7 | 32.87 | 22.71 | 9.0 | 36.6 |
| #3 | 0.96 | 0.96 | 0.96 | 0.10 | 0.02 | 10.3 | 43.8 | 15.1 | 14.8 | 15.1 | 4.8 | 1.1 | 3.2 | 13.6 | 66.4 | 78.2 | 66.7 | 57.2 | 5.1 | 2.5 | 13.5 | 262.4 | 263.2 | 263.2 | 26.9 | 36.69 | 26.33 | 10.6 | 42.7 |
| #4 | 0.94 | 0.95 | 0.95 | 0.10 | 0.02 | 10.2 | 43.1 | 18.9 | 18.6 | 18.9 | 5.1 | 1.2 | 3.7 | 15.9 | 52.9 | 58.5 | 53.1 | 20.9 | 3.5 | 3.2 | 15.8 | 307.9 | 309.5 | 309 | 30.4 | 40.81 | 29.47 | 11.4 | 45.6 |

No.=number, $T_2^*w$=$T_2^*$ weighted, Org.=original, w/=with, Den.=denoised, $R_2^*w$=$R_2^*$ weighted, ppb=parts per billion.

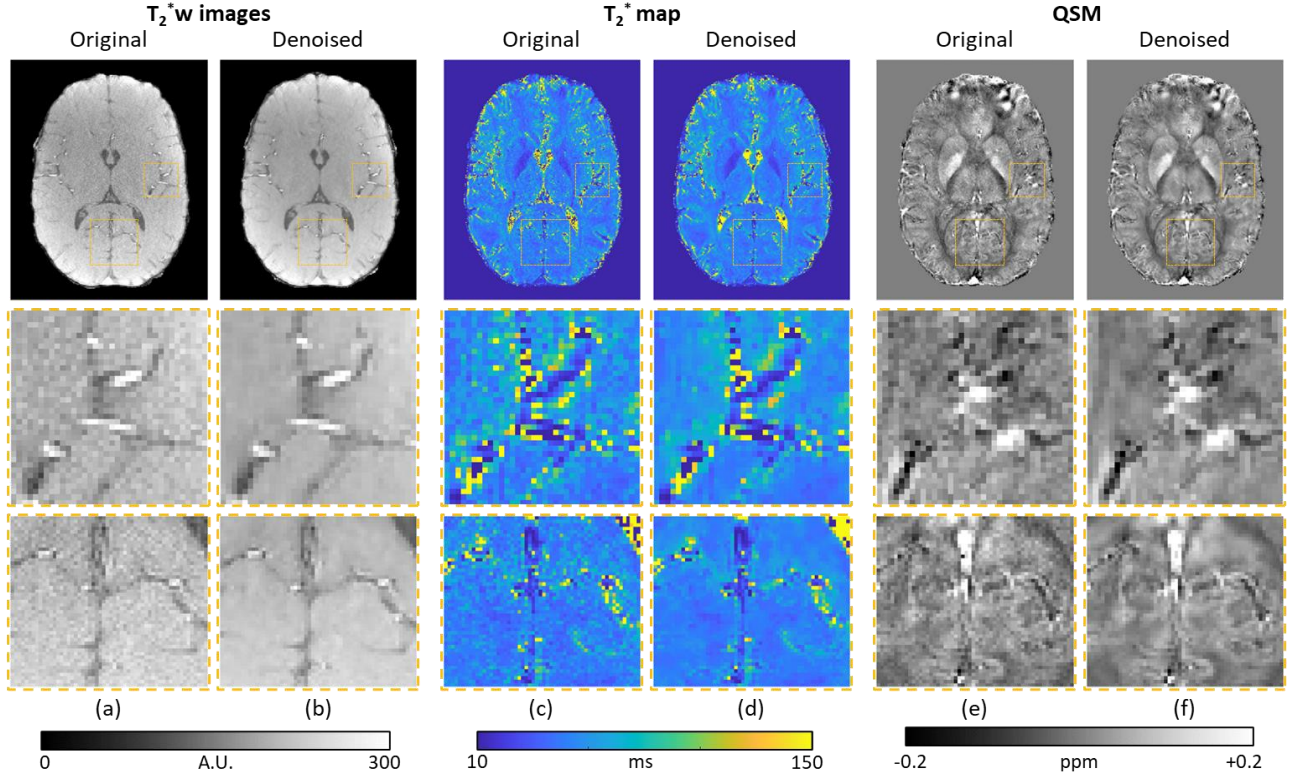

**Fig. 5.** MP-PCA denoising of brain anatomy of a healthy subject #4 pre- and post-denoising. (a-b) $T_2^*w$ images (first echo), (c-d) $T_2^*$ maps, (e-f) QSM maps pre- and post- denoising. Zoomed regions of interest are shown below each type of image/map (marked in orange dashed rectangles). Efficient denoising is achieved for all data types without visible loss of information.

An average decrease of 2.9 ± 0.9 % in $T_2^*$ values was observed post denoising, with a corresponding decrease of 23.9 ± 10.2 % in SD. The denoising pipeline did not induce any consistent trend in the QSM values, producing an average change of 0.5 ± 2.5 % in mean values and an average decrease of 10.8 ± 5.1 % in SD post-denoising. SNR in all assayed ROIs was estimated for $T_2^*w$ images, using four regions in the image background to represent noise magnitude (see section III). Consistent increase in SNR was observed for all ROIs with an average improvement of 74.2 ± 0.2 % post-denoising. The SNR values were also calculated for five more scans acquired with different scan parameters. For scan number 3, improvement of 139.2 ± 0.4 % post-denoising was observed. For scan number 4 which has the same acquisition parameters except acceleration factor of factor 2, the denoising process demonstrated an improvement of 139.3 ± 1.3 %. For scans 5, 6, and 7 we calculated improvement of 179.9 ± 0.8 %, 28.8 ± 0.3 %, and 252.8 ± 2.2 % respectively. These results demonstrate the impact of acquisition resolution, acquisition bandwidth, and acceleration factor on the SNR of the original and denoised images.

Fig. 6 depicts the QSM values for two subjects with brain metastasis pre- and post-denoising. Zoomed in view of the tumor areas indicates that the denoising process preserved the anatomical details of the tissue while improving the image's SNR. As described in section II, the inverse problem in QSM mapping is ill-posed, leading to potential amplifications of noise during its reconstruction. This can result in the introduction of streaking artifacts in the QSM maps, an issue highlighted in various studies [53], [54]. Such artifacts are evident in Fig. 7. Utility of MP-PCA denoising for decreasing QSM map artifact, caused by pulsating blood flow, and manifesting as a vertical line along the phase-encoding dimension (top to bottom). Figure, showing the additional utility of the denoising process in alleviating the streaking pattern in a patient with sickle cell disease.

## V. Discussion

This study evaluated MP-PCA denoising of $T_2^*w$ images and derived $R_2^*$, $T_2^*$ and QSM maps. Successful application of the technique was demonstrated on numerical phantom where both the accuracy and precision of all measured values were increased, showing robustness to different noise patterns. In vivo validations were also performed for both healthy and

TABLE III
SNR VALUES FOR $T_2^*$ WEIGHTED IMAGES, ALONGSIDE QUANTITATIVE $R_2^*$, $T_2^*$, AND QSM VALUES PRE- AND POST-DENOISING FOR SIX ROIs, SEGMENTED USING FREESURFER FOR A HEALTHY VOLUNTEER.

| | $R_2^*$ map | | | | | | $T_2^*$ map | | | | | | QSM | | | | | | SNR $T_2^*$w | | |
|---|---|---|---|---|---|---|---|---|---|---|---|---|---|---|---|---|---|---|---|---|---|
| | Mean [$s^{-1}$] | | | SD [$s^{-1}$] | | | Mean [ms] | | | SD [ms] | | | Mean [ppb] | | | SD [ppb] | | | | | |
| | Org. | Den. | Diff. [%] | Org. | Den. | Diff. [%] | Org. | Den. | Diff. [%] | Org. | Den. | Diff. [%] | Org. | Den. | Diff. [%] | Org. | Den. | Diff. [%] | Org. | Den. | Diff. [%] |
| L. G. Pallidus | 35.7 | 36.2 | 1.5 | 8.3 | 6.5 | -21.6 | 30.0 | 28.9 | -3.7 | 9.4 | 7.4 | -20.4 | 107.8 | 107.2 | -0.5 | 78.5 | 73.6 | -6.3 | 10.3 | 18.0 | 73.9 |
| R. G. Pallidus | 37.2 | 37.7 | 1.5 | 8.0 | 6.6 | -17.5 | 28.3 | 27.5 | -2.8 | 7.3 | 6.2 | -14.0 | 143.7 | 142.4 | -0.9 | 58.3 | 54.2 | -7.1 | 10.2 | 17.7 | 74.1 |
| L. C. Nucleus | 20.6 | 20.7 | 0.5 | 4.6 | 3.5 | -24.7 | 51.1 | 49.6 | -2.9 | 11.7 | 7.7 | -34.3 | 53.0 | 54.3 | 2.4 | 40.1 | 36.0 | -10.3 | 10.3 | 17.9 | 74.3 |
| R. C. Nucleus | 21.2 | 21.3 | 0.1 | 5.5 | 5.1 | -7.5 | 50.6 | 50.0 | -1.2 | 14.8 | 13.4 | -10.0 | 53.1 | 51.0 | -4.0 | 27.0 | 24.3 | -10.0 | 9.7 | 16.9 | 74.1 |
| L. Putamen | 23.4 | 23.8 | 1.7 | 4.6 | 3.2 | -30.5 | 44.4 | 42.7 | -3.8 | 9.3 | 5.7 | -38.0 | 44.9 | 46.1 | 2.8 | 28.6 | 22.4 | -21.6 | 10.8 | 18.8 | 74.4 |
| R. Putamen | 22.5 | 22.8 | 1.5 | 4.5 | 3.6 | -20.0 | 46.2 | 44.8 | -3.1 | 9.3 | 6.8 | -27.1 | 47.9 | 46.4 | -3.0 | 34.4 | 31.1 | -9.5 | 10.1 | 17.7 | 74.4 |

ROI=region of interest, L.=left, R.=Right, G. Pallidus=Globus Pallidus, C. Nucleus=Caudate Nucleus, Org.=original, Den.=denoised, Diff.=Difference, SD=standard deviation pathological brains, demonstrating effective denoising across various scan settings including acceleration factor, resolution, echo times, and acquisition bandwidth. Anatomical details were consistently preserved for all assayed settings with no visible loss of information. A notable advantage of the MP-PCA denoising algorithm as a versatile pre-processing tool, is that it requires no preliminary assumptions or adjustments of the hardware or scan parameters. In contrast to previous studies [28]–[30], this study showed that effective denoising can be achieved using a relatively small number of echoes. Moreover, the denoising process allowed to push the limits of the acquisition protocol to sub-millimeter resolutions up to 0.5 mm³ and low flip angles, without the typical tradeoff of SNR.

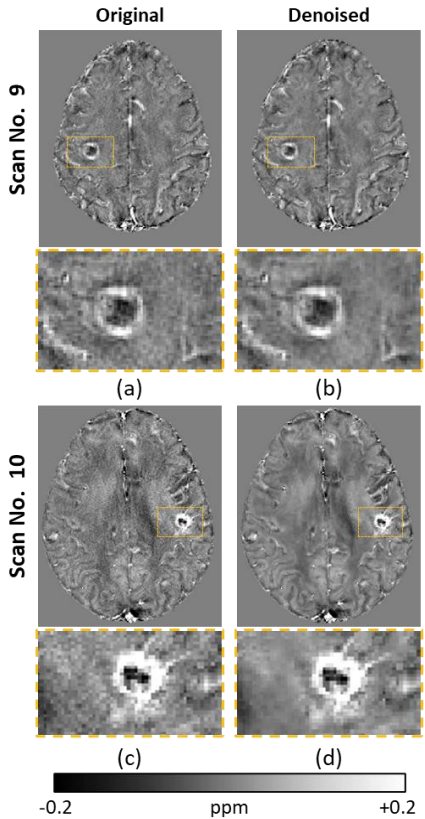

**Fig. 6.** MP-PCA denoising of two patients with brain metastasis, for the first patient (a-b) the lesion appears next to the postcentral gyrus, and for the second (c-d) it appears next to the precentral gyrus. (a, c) QSM of the original datasets, (b, d) QSM post-denoising datasets. Zoomed regions of interest are shown below each map (marked in orange dashed rectangles).

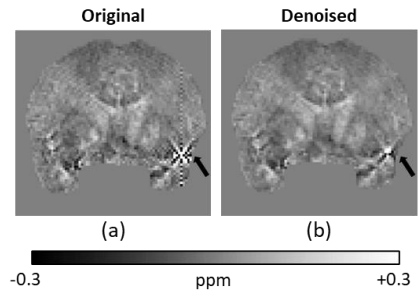

**Fig. 7.** Utility of MP-PCA denoising for decreasing QSM map artifact, caused by pulsating blood flow, and manifesting as a vertical line along the phase-encoding dimension (top to bottom). Figure presents coronal brain slice of a QSM map pre- and post- denoising in a patient with sickle cell anemia. (a) Artifacts are presented in the pre-denoised image, marked with a black arrow. (b) Post-denoising map shows a significantly attenuated artifact.

### A. Considerations of MP-PCA denoising

Several factors influence the denoising procedure, ultimately determining the final SNR levels and quantitative values. First, scanning parameters, mainly spatial-resolution and acquisition bandwidth will affect SNR. Using MP-PCA denoising, these limitations can be relaxed, allowing the use of higher resolution or higher acquisition bandwidth without trading off SNR. This is demonstrating how MP-PCA denoising successfully improved the SNR across various scan parameters and acceleration factors. Since all data were acquired using a 3D multi-GRE protocol, a 3D isotropic moving window was used for all datasets with the number of echoes determining the fourth dimension (see section II). A minimal window size of 2 × 2 × 2 pixels was found to be optimal for denoising, resulting in a 4D window of 2 × 2 × 2 × $N_E$, vectorized to $[N_V \times N_E] = 8 \times N_E$. Since the number of eigenvalues of PCA decomposition is determined by the smallest dimension (minimum of $N_V$ and $N_E$) [36], choosing a larger isotropic spatial window would not have changed the number of PCs in each window, seeing as multi-GRE protocols are limited in the number of echoes, typically 5-10 [34]. Following similar consideration, Does *et al.* [29] used a 2D window size of $N_W \times N_W$ (for a 2D acquisition), where $N_W \approx \sqrt{N_E}$, corresponding to our use of a 3D window size of $N_W \approx \sqrt[3]{N_E}$.

Parallel imaging, such as GRAPPA, is generally recommended for QSM [34], yet can lead to spatial dependency of the noise across the imaged FOV, which may adversely affect the performance of the denoising algorithm. Our findings reveal that despite this potential challenge, denoising performance was not impaired when using moderate x2 acceleration. This is described in section IV, where the relative elevation in SNR was similar between accelerated and non-accelerated data (scans 3 and 4).

### B. Noise patterns and propagation in $R_2^*$ and $T_2^*$ maps

The addition of noise in the simulation led to an elevation of $T_2^*w$ values compared to baseline. This was caused due to the non-linear combination of real and imaginary images, altering the noise distribution, a process extensively elaborated in [52]. As reported, for large SNR values (>3), noise pattern closely resembles a Gaussian distribution, retaining the original variance ($\sigma^2$) but with a mean value adjusted to $\sqrt{I_0^2 + \sigma^2}$, where $I_0$ is the original magnitude of the signal. As SNR decreases below 3, e.g., in later echo times, the noise distribution shifts to Rician, leading to an even more pronounced elevation of the signal magnitude, as demonstrated by Stern *et al.* [30]. These effects were mitigated post-denoising, allowing to use signal points which were previously below the noise level. It is important to emphasize that similar to [29], [30], [33], denoising was applied to complex $T_2^*w$ data (magnitude and phase) to maintain the Marchenko-Pastur noise distribution and minimize Rician noise at later echo times. The propagation of noise was further investigated when transitioning from $T_2^*w$ images to $R_2^*$ and $T_2^*$ maps. Analysis showed that pre-denoising, $R_2^*$ values exhibited a small and non-systematic deviation from original values, unlike the corresponding $T_2^*$ values which demonstrated a significant overestimation of values. This is likely attributed to the proximity of $R_2^*$ values to zero, leading to amplification of noise during the division operation used for calculating $T_2^*$ values. This particularly affected high $T_2^*$ values corresponding to low $R_2^*$ values (see TABLE II). We therefore recommend relying on $R_2^*$ values where possible to minimize augmentation of noise during post-processing.

### C. Denoising of QSM data

The variations in the main magnetic field $\Delta B_0(\vec{r})$ are at the core of QSM and are obtained from the phase images. Previous studies report that the noise at each voxel in the phase image has a Gaussian distribution, while its standard deviation is inversely proportional to the corresponding voxel's magnitude, which, in turn, will depend on the local proton density, and $B_1^+/B_1^-$ profiles [50], [51]. During the QSM pipeline, this noise propagates through several non-linear operations (see section II), which alter its Gaussian distribution and introduce further variability among voxels with the same susceptibility. Using the denoising pre-processing the phase noise is also reduced, thereby mitigating the non-linear variability induced by the QSM pipeline without loss of information as demonstrated in

Fig. 2. MP-PCA denoising of a numerical phantom. (a-d) $T_2^*w$ images (4th echo) pre- and post-denoising. (e-h) QSM maps pre- and post-denoising. Simulations were done at low SNRs of 10 & 20. (a) shows the location of tubes (1)-(4). and in TABLE II. The QSM pipeline implemented in our study followed the model outlined in [35] using the pipeline's default regularization parameters. The integration of MP-PCA denoising into other QSM pipelines or using regularization parameters is expected to be similarly effective and remains a prospect for future research. One example is the truncated K-space division (TKD) [1], a common method for a rapidly obtaining solution to the inverse problem in **Error! Reference source not found.**). The TKD technique is highly sensitive to noise [53] leading to potentially significant artifacts in the susceptibility map, making it an ideal candidate for denoising algorithms.

QSM is valuable for differentiating intratumoral hemorrhage in metastatic brain tumors, from calcifications that may result from therapeutic response. Fig. 6, show differences between two patients with brain metastases, where the first patient's QSM map contains a lesion with pronounced calcification (potentially indicative of treatment efficacy [55]), and the second patient's QSM map exhibits a lesion with much thicker rim associated with prominent hemorrhagic pathology. Improving the diagnostic quality of QSM can thus be useful in the clinic, allowing more precise comparison between metastases from different origin, treatment procedures, and for comparing different time points in longitudinal studies [56].

## VI. CONCLUSIONS

The current study highlights the promising potential of MP-PCA denoising for enhancing the diagnostic quality of QSM maps by not only improving SNR but also alleviating image artifacts and enabling the acquisition of higher resolution data. Recognizing the need for larger test cohort, these initial results nevertheless provide a sound proof-of-concept for the utility of MP-PCA denoising of $T_2^*$ weighted time series and their quantitative $T_2^*$, $R_2^*$ and QSM derivatives.

## VII. APPENDIX

To assess the impact of different coil combination schemes, MP-PCA denoising was applied on data acquired without acceleration for one healthy volunteer using three different coil combination procedures: adaptive combination done directly on the scanner; adaptive combination done during post-processing based on the method described by Bernstein *et al.* [63]; similar adaptive combination but done after the denoising process, which was implemented for each of the 16 channels separately. $T_2^*$ and QSM maps were then compared for each coil combination scheme.

A comparison between denoising each coil $T_2^*w$ data pre- and post- Berenstein's coil combination [63] is presented in Fig. 8. It can be seen that the denoising not only improved image quality but also reduced artifacts from phase noise.

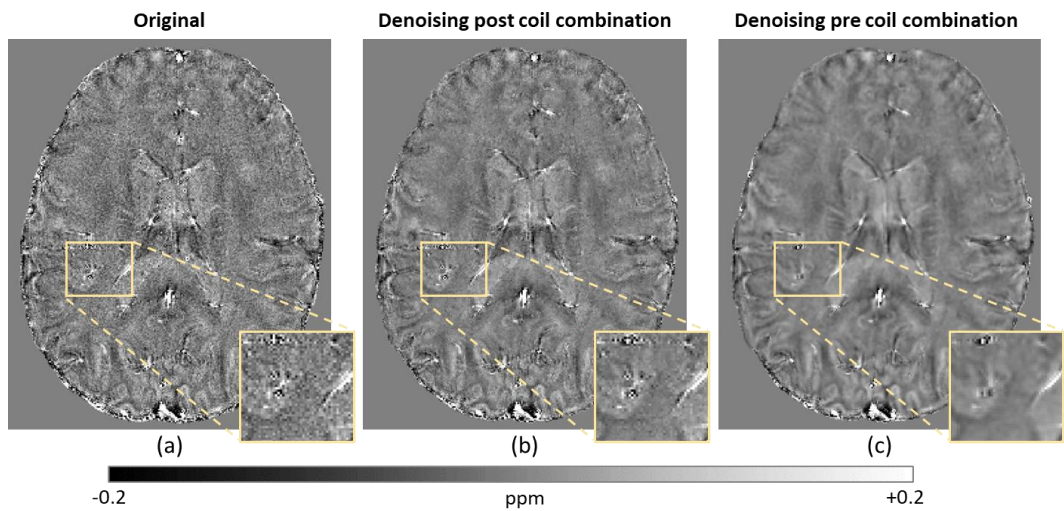

**Fig. 8.** QSM maps of an In vivo subject 1, and voxel size of an isotropic 0.6 mm³. **(a)** QSM based on the original data. **(b)** QSM that was derived from the denoising of the complex $T_2^*w$ coil combined data. **(c)** QSM was derived by denoising each channel of the complex $T_2^*w$ data and then coil combined.

The denoising algorithm showed improvements in all three cases. However, processing raw data from each channel separately yielded the most significant visual enhancements. A notable drawback of working with raw data is the increased processing time due to denoising each channel individually. Another limitation is the typical unavailability of raw data, particularly for retrospective analyses, or the necessity of manual intervention. Given these limitations, and following the recommendations in the QSM Consensus paper [44], our study predominantly utilized DICOM files.

## VIII. ACKNOWLEDGMENTS

Karin Shmueli and Mitchel Lee were supported by European Research Council Consolidator Grant DiSCo MRI SFN 770939.

I acknowledge the use of ChatGPT 4 (Open AI, https://chat.openai.com) to summarize my initial notes and to proofread my final draft.